\documentclass[fleqn,10pt,dvipsnames]{wlscirep}
\usepackage[utf8]{inputenc}
\usepackage[T1]{fontenc}
\usepackage{amsmath,graphicx}
\usepackage{color}
\usepackage{subfig}
\usepackage[modulo]{lineno}
\usepackage{fonttable}
\usepackage{newtxtext,newtxmath}
\usepackage{pgfplots}
\usepackage{tikz}
\usepackage{import}
\usepackage{adjustbox}      

\newcommand{\red}[1]{\textcolor{black}{#1}}
\newcommand{\redrev}[1]{\textcolor{black}{#1}}

\newcommand{\vect}[1]{\boldsymbol{#1}}
\newcommand{\imdimthree}[3]{{#1}$\,\times\,${#2}$\,\times\,${#3}}

\title{Projection-to-Projection Translation for Hybrid X-ray and Magnetic Resonance Imaging}

\author[1,2,*]{Bernhard~Stimpel}
\author[1,2]{Christopher~Syben}
\author[1]{Tobias~W\"urfl}
\author[1]{Katharina~Breininger}
\author[2]{Philip~Hoelter}
\author[2]{Arnd~D\"orfler}
\author[1]{Andreas~Maier}

\affil[1]{Pattern Recognition Lab, Friedrich-Alexander University Erlangen-Nürnberg, Erlangen, Germany}
\affil[2]{Department of Neuroradiology, University Hospital Erlangen, Erlangen, Germany}
\affil[*]{bernhard.stimpel@fau.de}

\begin{abstract}
Hybrid X-ray and magnetic resonance (MR) imaging promises large potential in interventional medical imaging applications due to the broad variety of contrast of MRI combined with fast imaging of X-ray-based modalities. To fully utilize the potential of the vast amount of existing image enhancement techniques, the corresponding information from both modalities must be present in the same domain. For image-guided interventional procedures, X-ray fluoroscopy has proven to be the modality of choice. Synthesizing one modality from another in this case is an ill-posed problem due to ambiguous signal and overlapping structures in projective geometry. To take on these challenges, we present a learning-based solution to MR to X-ray projection-to-projection translation. 
We propose an image generator network that focuses on high representation capacity in higher resolution layers to allow for accurate synthesis of fine details in the projection images.
Additionally, a weighting scheme in the loss computation that favors high-frequency structures is proposed to focus on the important details and contours in projection imaging. 
The proposed extensions prove valuable in generating X-ray projection images with natural appearance. 
Our approach achieves a deviation from the ground truth of only $6$\,\% and structural similarity measure of $0.913\,\pm\,0.005$. 
In particular the high frequency weighting assists in generating projection images with sharp appearance and reduces erroneously synthesized fine details. 
\end{abstract}

\begin{document}
\flushbottom
\maketitle
\thispagestyle{empty}
%\noindent Please note: Abbreviations should be introduced at the first mention in the main text – no abbreviations lists. Suggested structure of main text (not enforced) is provided below.

%\linenumbers

\section{Introduction}
\label{sec:introduction}
Medical imaging offers various possibilities to visualize soft and hard tissue, physiological processes, and many more. The range of information that can be acquired is, however, divided between many different modalities. Hybrid imaging has the potential to provide simultaneous access to distributed information for diagnostic and interventional applications~\cite{Fahrig2001,Wang2013, Wang, Gjesteby2016}. For example, future research advances can use the combination of computed tomography~(CT) and magnetic resonance imaging~(MRI) for clinical applications. This is of particular interest for interventional applications, as a large number of tasks are associated with the manipulation of soft tissue. Despite this fact, X-ray imaging, which is insensitive to soft tissue contrast, is still the workhorse of interventional radiology. The reasons for this are the high spatial and temporal resolution, which make the handling of interventional devices much easier. By complementing these advantages with better soft tissue contrast provided by MRI, the gain from the simultaneous acquisition of soft and dense tissue information through hybrid imaging would offer great opportunities. 

Assuming that the information from both modalities is available simultaneously, numerous existing post-processing methods become applicable. Image fusion techniques, such as image overlays, have proven successful in the past. Additionally, techniques for image enhancement, such as noise reduction or super-resolution, can be considered. 
For many of the previously presented methods it is advantageous to have the data available in the same domain. 
The generation of CT images from corresponding MRI data was previously presented~\cite{Navalpakkam2013, Nie2017, Wolterink2017,Xiang2018}, mainly to create attenuation maps for radiation therapy. However, all are applied to volumetric data, i.e., to tomographic images.

Contrarily, interventional radiology is strongly dependent on line-integral data originating from projection imaging. Images with similar perspective distortion can be acquired directly with an MR device~\cite{Lommen2018a, Syben2018}. This avoids time-consuming volumetric MRI acquisition and subsequent forward projection. 
The synthesis of X-ray-like projection images for further processing based on the acquired MRI signal is, however, an inherently ill-posed problem. 
The X-ray imaging signal is dominated by dense tissue structures, e.g., bone, which provide almost no signal in MRI. 
As air also does not induce signal in MRI, the materials cannot be discriminated based on intensity values alone.
Resolving this ambiguity is, therefore, only possible based on the available structural information. 
In volumetric images, the materials may be unknown but are resolved in distinct regions of the image.
While MR projection imaging allows for continuous intraoperative use due to much faster acquisition times compared to volumetric MRI scans, the structural information diminishes in the projection image by integrating the intensity or attenuation values on the detector.
This corresponds to a linear combination of multiple slice images with unknown path length which further increases the difficulty of the synthesis task. 
%To the best of our knowledge, no approach to tackle this problem has been proposed up to now. 
Enabled by the progress in fast MR projection acquisition, we investigate a solution for generating X-ray projections from corresponding MRI views by projection-to-projection translation.

\section{Related Work}
\label{sec:related_work}
\subsection{MR Projection Imaging}
\label{sec:related_work_mrpi}
Common clinical X-ray systems used for image-guided interventions exhibit a cone-beam geometry with the according perspective distortion in the acquired images. 
These systems can acquire fluoroscopic sequences with a rate of up to 30 frames per second. In contrast, MR imaging is usually performed in a tomographic fashion. The acquisition of whole volumes and subsequent forward projection is, however, too slow to be applicable in interventional procedures. 
Fortunately, the direct acquisition of projection images is possible~\cite{Lommen2018a, Syben2018}. Yet, the resulting images are subject to a parallel-beam geometry and, therefore, incompatible with their cone-beam projected X-ray counterpart. 
To remedy this contradiction, recent research has addressed the problem of acquiring MR projections with the same perspective distortion as an X-ray system, without the detour of volumetric acquisition~\cite{Wachowicz2018}. Most approaches rely on rebinning to convert the acquired projection rays from parallel to fan- or cone-beam geometry~\cite{Syben2017, Lommen2018a}. 
Unfortunately, this requires interpolation which reduces the resulting image's quality.  To avoid loss of resolution, Syben et al.~\cite{Syben2018} proposed a neural network based algorithm to generate this perspective projection. 

\subsection{Image-to-Image Translation}
Projection-to-projection translation suffers from the problems explained in Section \ref{sec:introduction}, i.e., ambiguous signal and overlapping structures. However, from a machine learning point of view, it is essentially an image-to-image translation task. 
While we target the application of synthesizing 2D projection images instead of tomographic slice images, the preliminary work on pseudo-CT generation, mostly applied for radiotherapy treatment planning, is still relevant. Two main approaches to the task of estimating 3D CT images from corresponding MR scans have prevailed up to now: atlas-based and learning-based methods. Atlas-based methods as proposed in~\cite{Uh2014,Degen} achieve good results, however, their dependence on accurate image registration and often high computational complexity is undesirable in the interventional setting. 
In contrast, inference is fast in learning-based approaches which makes them suitable for real-time applications. Essentially a regression task, image-to-image translation can be solved using classical machine learning methods like random forests~\cite{Navalpakkam2013}. The advent of convolutional neural networks~(CNN) revolutionized the field of image synthesis and image processing in general. Numerous different approaches to image synthesis using deep neural networks have been published since then~\cite{Yan2015,Zhang,Chen2017}. Lately, generative adversarial networks~(GAN)~\cite{Goodfellow2014}  proved valuable for this task. Isola et al.~\cite{Isola2016}  presented one of the first approaches to general-purpose image-to-image translation based on GANs and others followed~\cite{Nie2017,Wolterink2017,NIPS2017_6672,Zhu2017,Wanga,Huang_2018_ECCV,Choi_2018_CVPR}. 

Applying this idea to the field of medical image synthesis, Nie et al. enhanced the conditional GAN structure with an auto-context model and achieved good results in the task of tomographic CT image synthesis based on its MR pendant~\cite{Nie2017}. Furthermore, in~\cite{Yang2018, Wolterink2017, 10.1007/978-3-030-00536-8_4,chartsias2017adversarial}, the successful training of a GAN based on unpaired MR and CT images was shown. 
\redrev{As a result of the achieved successes, GANs are now a frequently used tool in medical imaging research that extends beyond the realm of image-to-image translation~\cite{10.1007/978-3-030-00536-8_1,Armanious2018,Yang2018a, Yu2019, yi2019generative}.}
Inspired by these advances in the field of image synthesis, we seek to find a deep learning-based solution to generate X-ray projections from corresponding MR projections. In previous research \cite{Stimpel2019a}, we were able to show a proof of concept, which however was only able to produce mediocre images. We want to remedy this shortcoming in the underlying work and additionally show a profound analysis of the results on a more representative data set.

\section{Methods}
\label{methods}

\begin{figure}
	\centering
	\includegraphics[width=1\textwidth]{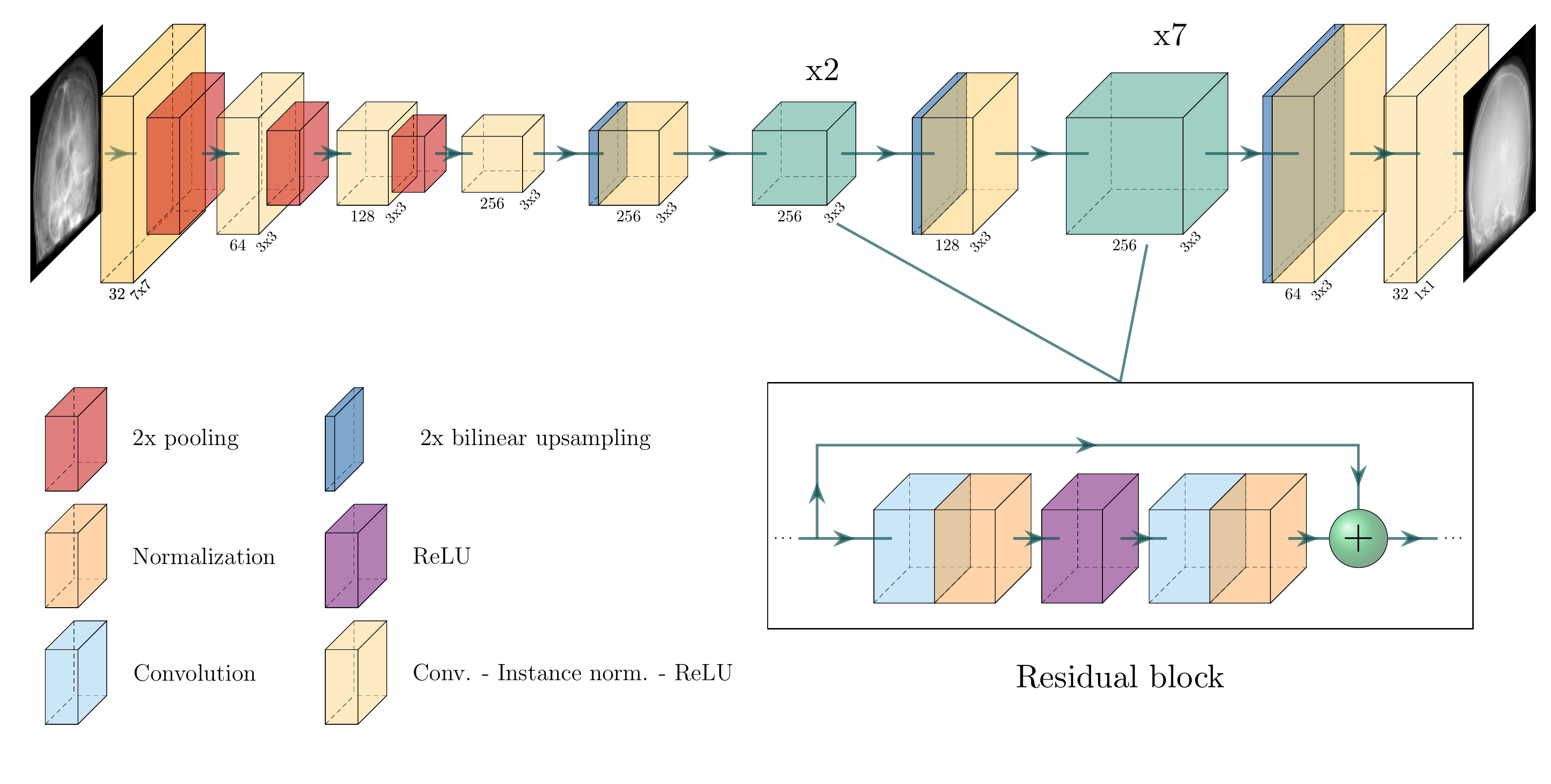}
	\caption{\red{A schematic visualization of the proposed network architecture. The numbers on the layers denote the feature dimensions and the convolution kernel size.} }
	\label{fig:network_graph}
\end{figure}

In contrast to most natural image synthesis problems we do not desire to learn a one-to-many mapping. While there are different "correct" solutions for synthesized images from, e.g., a semantic layout, only one solution is correct for projection-to-projection translation tasks. The underlying network can, therefore, be trained in a supervised manner based on corresponding pairs of MR and X-ray projection images. Please note that the mapping itself remains an ill-posed problem due to the previously discussed signal ambiguities and structural overlaps. Formally, we seek to learn a mapping that generates an image $\vect{G}$ based on an input image $\vect{I}$. The mapping is trained based on corresponding input and label image pairs, $\vect{I}$ and $\vect{L}$, such that the generated image $\vect{G}$ is as close as possible to $\vect{L}$. In the underlying task, $\vect{I}$ and $\vect{L}$ are the MR and X-ray projection images, respectively, and $\vect{G}$ is the generated X-ray projection.  
\subsection{Network architecture}
\label{sec:methods_architecture}
We employ a fully convolutional neural network as our image generator. The general considerations regarding the network's architecture are based on the popular approaches proposed by~\cite{Johnson2016,Wanga,Zhu2017,Wolterink2017} which have shown promising results also in medical imaging applications~\cite{Wolterink2017}. The network's architecture is designed in an encoder-decoder fashion. In the originally proposed approaches, the lowest resolution level consists of a series of subsequent residual blocks~\cite{He2015}. These blocks allocate the majority of network capacity that is utilized for the projection-to-projection translation. At the lowest, most coarse resolution layer the general outline of structures and their alignment in the image are determined. Considering the overwhelming variety of possible solutions that exist in the synthesis of natural images, the use of a large part of the available resources for the basic determination of the images seems reasonable. In contrast, the underlying variance observed in medical projection imaging is limited. In addition, the most insightful information during interventional procedures is often related to outlines of bones and organs, medical devices, and similiar structures. These are represented by high-frequency details in the image in the form of edges and contrasts. However, sharp borders and edges are synthesized at the high-resolution layers of the network. To increase the network capacity at the higher resolution stages, we redistribute the residual blocks to these layers. 
Since the memory requirements increase with increasing resolution, the placement on higher layers must be weighted against a decreased amount of feature channels. 
In Figure \ref{fig:network_graph} a schematic visualization of the architecture is given.

\subsection{Objective Function}
\label{methods:objfunc}
The appearance of the generated images is highly dependent on the generator's objective function. In contrast to natural image synthesis tasks, only one unique solution is valid for the underlying projection-to-projection translation problem. Therefore, computing intensity-based loss-functions is suitable. Yet, the results favored by a pixel-wise loss do not necessarily correspond to perceived visual quality~\cite{Dosovitskiy2016}. Image generators trained based on these metrics tend to produce images that are far too smooth. To avoid this, we employ a perceptual objective function for this approach. 
The main component of the proposed objective function is an adversarial loss scheme as proposed by~\cite{Goodfellow2014}. A discriminator network is trained to tell apart fake generated images from real label images and provide the generating network with a gradient. We adopt the architecture proposed in~\cite{Zhu2017} for our discriminator. Although the adversarial loss offers powerful guidance, it is also less constrained to the real target image than common objective functions. Therefore, it is often combined with one or more additional metrics. To provide additional high-level guidance while simultaneously limiting the deviations from the target image, we add a feature matching loss to the objective function~\cite{Gatys2016}. To this end, the generated and label image are fed through a fixed, pre-trained network. The loss is computed by comparing the feature activations of both images which are subject to a surjective, i.e., non-unique mapping of the images to the feature space.
If both feature activations are equal, the respective images are equal w.r.t. the mapping, too. Increasing deviation in the feature activations is a strong indicator for deviating images and, consequently, the error increases. 

As initially motivated, pixel-wise metrics are usually suitable for one-to-one mappings as it is the case here. The projection images are, however, dominated by homogeneous regions. This is problematic, as the loss induced by low-frequency structures may overwhelm the error in the high-frequency parts. For X-ray fluoroscopy, these form most of the valuable information in the image that is perceived by the physicians. To alleviate this problem and put further emphasis on the correct creation of high-frequency detail, we propose to include a high-frequency weighting to the loss computation. 
Accordingly, a weighting map has to be generated that describes how important each pixel is for the overall impression. 

First, the Sobel filter~\cite{Sobel1968} is used to compute the gradient map of the label image. The resulting gradient maps correspond to the likelihood of a pixel belonging to an edge. High likelihood is represented by high intensity values and vice versa. Representative examples are presented in Fig. \ref{fig:projection_results}. Second, this gradient map is used to weight the loss such that the loss generated from edges is emphasized and that from homogeneous regions is attenuated. The effectiveness of including this edge information in image synthesis tasks has previously been shown for projection images in~\cite{Stimpel2019a} and recently also for other medical imaging domains~\cite{Yu2019}. Starting with the GAN loss, this can be formulated as
\begin{equation}
\label{eq:gan_loss}
\ell_\text{GAN}(\vect{L},\vect{G}, \mathcal{D}) = \mathbb{E}_{\vect{L},\vect{G}}\left[	\log \mathcal{D}(\vect{L},\vect{G})\right] +  \mathbb{E}_{\vect{L},\vect{G}}\left[1-	\log \mathcal{D}(\vect{G})\right]
\end{equation}
where $\mathcal{D}$ is the discriminator network and  $\vect{L}$ and  $\vect{G}$ are the label and generated image, respectively. The second part of the loss function is the feature matching loss described by
\begin{equation}
\label{eq:fm_loss}
\ell_\text{FM}(\vect{L},\vect{G}) = \sum_{s}^{S} \left( \vect{V}_s(\vect{L}) - \vect{V}_s(\vect{G}) \right) \;,
\end{equation}
where $\vect{V}_s(\vect{L})$ and $\vect{V}_s(\vect{G})$ are the feature activation maps of the VGG-19 network~\cite{Simonyan2015} at the layer $s \in S$. This leads to the final objective functions
\begin{equation}
\label{eq:edge_loss}
\ell(\vect{L},\vect{G},\mathcal{D}) = \left(\ell_\text{GAN}(\vect{L},\vect{G}, \mathcal{D}) + \ell_\text{FM}(\vect{L},\vect{G})  \right) \cdot \vect{E_L}\;,
\end{equation}
which is the combination of the GAN and feature matching loss weighted by the gradient map $\vect{E_L}$ of the label image. For a simplified representation, we assumed here that all elements have the same dimensions. In practice, the outputs of the feature matching and GAN loss are multi- and low-dimensional, respectively, which is why the precalculated edge map has to be adjusted. We use bilinear downsampling for this.

\section{Experiments}
\label{experiments}
\paragraph{\redrev{Data Generation}}
\label{experiments_data_generation}
\begin{table*}[h]
	\caption{\red{A selection of the important properties of the data used for the underlying work.}}
	\centering
	\begin{tabular}{lcc}
		& CT data & MR data   \\
		\hline
		Type & unenhanced spiral & time-of-flight angiography \\
		& h70 (''sharp'') reconstruction kernel & \\
		Dimensions & \imdimthree{512}{512}{(120-368)} & \imdimthree{(256-412)}{(256-512)}{(176-188)} \\
		Spacing & \imdimthree{(0.39-0.45)}{(0.39-0.45)}{(0.6-1.2)}\;\text{mm}$^3$ & \imdimthree{0.39}{0.39}{0.5}\;\text{mm}$^3$
	\end{tabular}
	\label{tab:data_properties}	
\end{table*}
\color{black}
Evaluating the performance based on real-world data is crucial. However, since the acquisition of corresponding projection images is not a part of the clinical routine up to now, these have to be artificially created. To this end, MRI and CT head datasets of 13 individuals with different pathologies are provided by the Department of Neuroradiology, University Hospital Erlangen, Germany. (MR: 1.5\,T MAGNETOM Aera / CT: SOMATON Definition, Siemens Healthineers, Erlangen / Forchheim, Germany). Tab. \ref{tab:data_properties} shows important details about the type and properties of the datasets. \red{However, please note that no special imaging or selection of data is performed for this work. All datasets were acquired as a part of routine clinical imaging procedures.}

\begin{figure}
	\centering
	\includegraphics[width=0.8\textwidth]{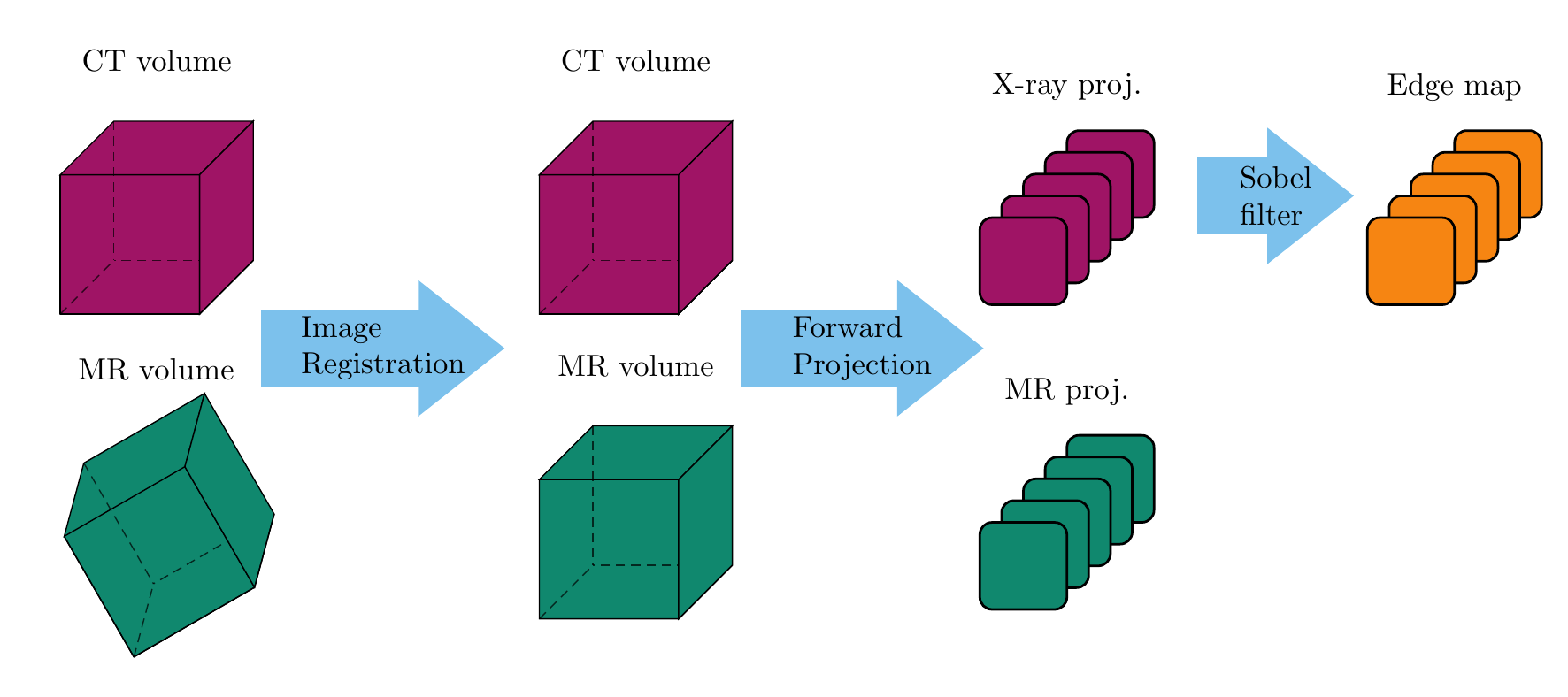} %
	\caption{\red{A schematic overview of the data generation process. Note that ultimately the goal is to acquire simultaneous projection images in hybrid X-ray and MR imaging (cf.~Section~\ref{sec:related_work_mrpi}). \redrev{To generate training data for the underlying work, simulation was necessary.}}}	
	\label{fig:data_gen}
\end{figure}

Rigid registration of the tomographic 3D data is performed using 3D Slicer~\cite{Fedorov2012}. Subsequently, the registered datasets are forward projected using the CONRAD framework~\cite{Maier2013} to create matching cone-beam MR and X-ray projection images. \redrev{The projection geometry and according parameters are set to closely match clinical C-arm CT systems~\cite{zeego}.} To ensure substantial variation while training, 450 projections per patient are created that are distributed equiangularly along the azimuthal and inclination angle. \redrev{During testing, projections from a $360^{\circ}$ rotation in the transversal plane are used which corresponds to common clinical acquisition procedures.} \red{In Fig. \ref{fig:data_gen}, a visual representation of the data generation process is given.} 

All projections are normalized prior to training by zero-centering and division by standard deviation. For the input data, i.e., the MRI projections, this preprocessing is applied to each subject's data individually and not on the whole dataset to account for differences in the MR protocols. Training is performed for a fixed number of 300 epochs using the ADAM optimizer~\cite{Kingma2014} with a learning rate of $1e^{-4}$ and a batch size of one. The edge map generated by the Sobel filter is thresholded with a value of 0.4 (data range 0 to 1) in order to abandon pixels with low likelihood of belonging to an edge. 
\paragraph{\redrev{Evaluation}}
\label{experiments_evaluation}
Evaluation of the proposed approach is performed quantitatively as well as qualitatively based on two patient datasets that were reserved for testing. Quantitative evaluation is done based on the mean absolute error (MAE), structural similarity index (SSIM), and peak signal-to-noise ratio (PSNR). In addition, the performance of these metrics with respect to the projection angle is investigated. All projections are normalized beforehand. For MAE and PSNR, only pixels are considered that are nonzero in the label images to limit the optimistic bias caused by the large homogeneous air regions.
For treatment planning in radiotherapy, a precise estimation of pixel-wise attenuation values is crucial. In interventional projection imaging, however, also the visual appearance of the generated images is of interest for the surgeon. Pixel-wise error metrics prefer smooth images, as explained previously in Section \ref{methods:objfunc}. Therefore, qualitative evaluation of the generated X-ray projection images is performed to make sure that the generated and reference images also correspond visually for the human observer. 

The proposed modifications of our approach are, first, a relocation of the residual blocks to higher resolution layers to increase the network capacity at the relevant stages and, second, a high-frequency weighting of the computed loss term.

We compare our approach to the conditional GAN architecture as used in~\cite{Johnson2016,Wanga,Zhu2017,Wolterink2017}, which we will refer to as reference architecture. The reference is similar to our proposed architecture (see Fig. \ref{fig:network_graph}) with the main difference that all residual blocks are placed at the lowest resolution layers. 
To ensure comparability, the network parameters of both approaches are chosen such that all available graphics card memory is utilized. 
\section{Results}
\label{sec:results}
\begin{figure*}
	\centering
	\includegraphics[width=1\textwidth]{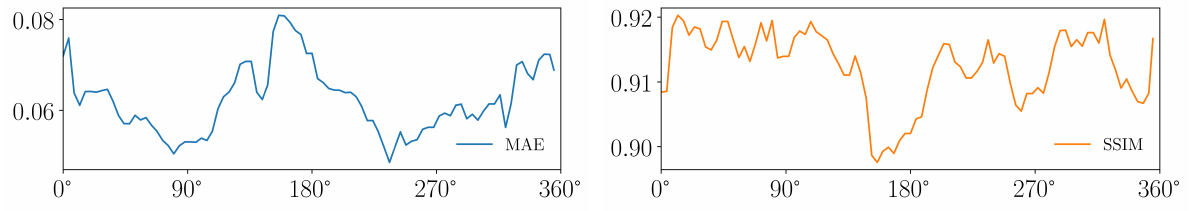} \hfil%
	\caption{Evaluation metrics with respect to the projection angle.}
	\label{fig:eval_angle}
\end{figure*}

\begin{figure*}
	\centering	
	\includegraphics[width=1\textwidth]{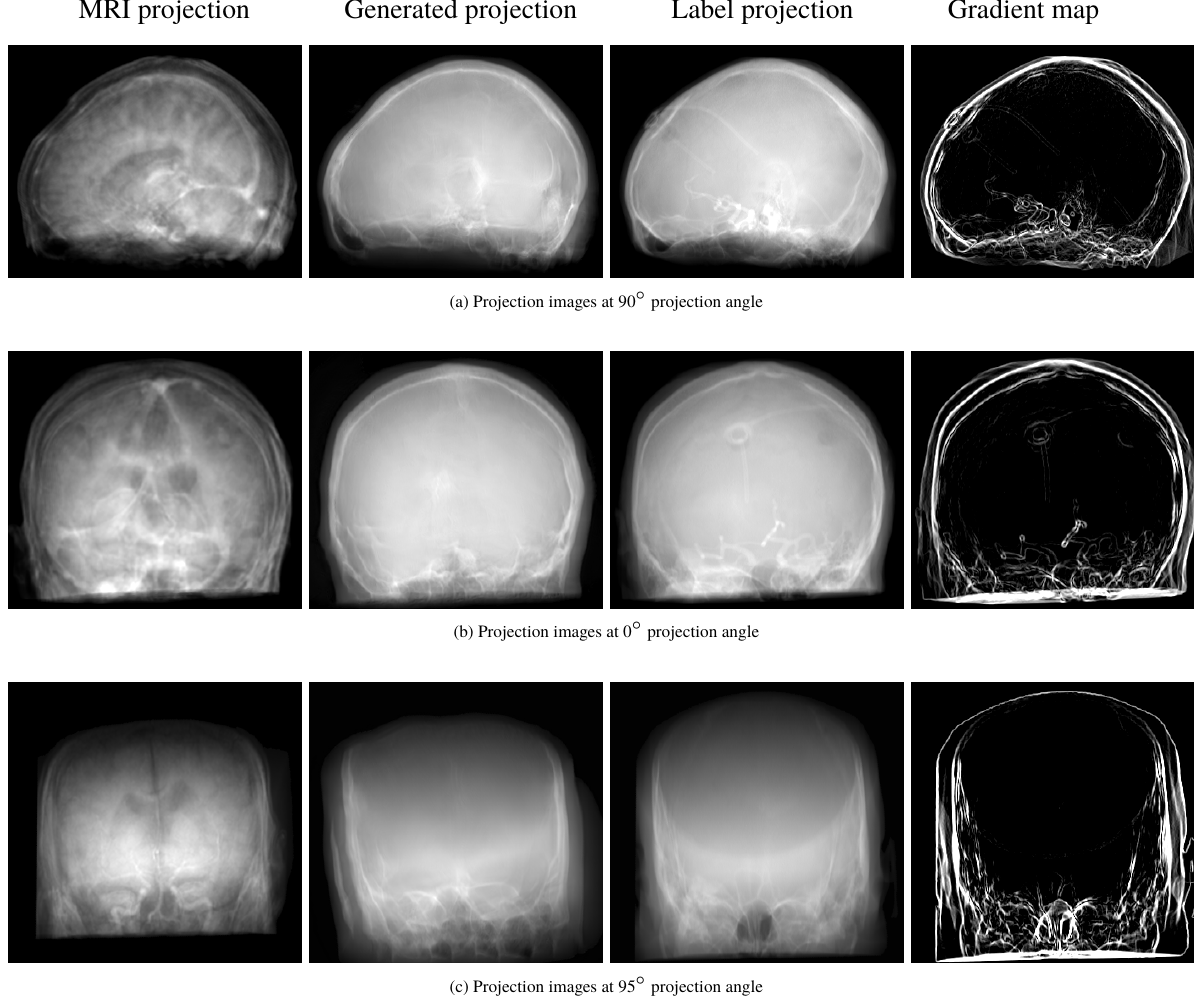} \hfil%
	\caption{Representative examples of the projection-to-projection translation for different projection angles and patients. \red{The top and middle row are projections originating from the first test patient and the bottom row from the second test patient.}
	}
	\label{fig:projection_results}
\end{figure*}

\begin{figure*}
	\centering
	\includegraphics[width=1\textwidth]{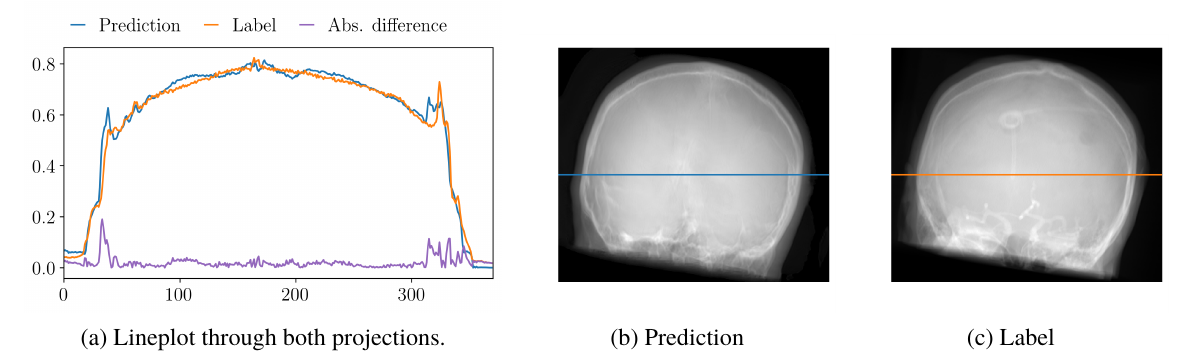}
	\caption{Exemplary lineplots of a pair of generated and label projection images.}
	\label{fig:lineplots}
\end{figure*}

\begin{figure*}
	\centering
	\includegraphics[width=1\textwidth]{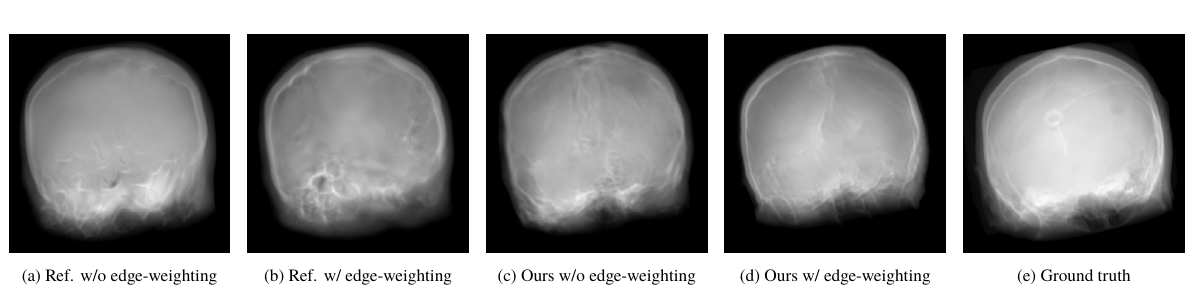}
	\caption{Influence of reference (ref.) and proposed architecture and loss functions on the generated results. }
	\label{fig:eval_architecture}
\end{figure*}

\begin{table*}[]
	\caption{MAE, SSIM, and PSNR of the different network architectures.}
	\centering
	\begin{tabular*}{0.71\linewidth}{l@{\extracolsep{0.2cm}}rrrr}
		\multicolumn{1}{l}{} & 
		\multicolumn{1}{c}{MAE in [\%]} & 
		\multicolumn{1}{c}{SSIM} & 
		\multicolumn{1}{c}{PSNR} & 
		\\ \hline
		Reference w/o edge-weighting	&	$7.7\,\pm\,1.7$			& 	$0.884 \,\pm\,0.011$				& 	$20.132\,\pm\,1.644$	&   \\ 
		Reference w/ edge-weighting	&	$6.6 \,\pm\,1.5$			& 	$0.902 \,\pm\,0.013$ & 	$21.558\,\pm\,1.802$	&   \\ 
		Ours w/o edge-weighting	&	$7.4\,\pm\,3.4$ 		& 	$0.909\,\pm\,0.011$				&	$21.480\,\pm\,3.187$ 	&	\\ 
		Ours w/ edge-weighting	&	$\mathbf{5.7\,\pm\,0.9}$& 	$\mathbf{0.913\,\pm\,0.005}$	&	$\mathbf{21.859\,\pm\,0.676}$ 	&	\\
	\end{tabular*}
	\label{tab:eval_networks}	
\end{table*}
\red{All quantitative metrics and observations reported in the following are computed or made based on the two representative test datasets.} In Tab. \ref{tab:eval_networks}, the quantitative results of the projection-to-projection translation are presented. We evaluated the proposed changes in architecture and the edge-weighting against the reference architecture. Shifting the residual blocks shows a small improvement in MAE as well as SSIM.
The combination of the proposed architecture and the edge-weighted loss function yields an improvement of 25\,\% compared to the reference architecture without the weighted loss term.
The gain in SSIM is not as large but still clear. In Fig. \ref{fig:eval_angle}, the error metrics are plotted w.r.t. the projection angle. In this figure, $0^\circ$ denotes an RAO angle of $90^\circ$ and, consequently, $180^\circ$  represents $90^\circ$ LAO. Note that opposing projections are not equal due to tilt and the projective geometry. 
The qualitative examples give a good intuition of the effects of the proposed changes. Representative examples of the generated images are given in Fig. \ref{fig:projection_results} for qualitative examination. The influence of the proposed changes regarding the network architecture and the edge-weighting can be observed in Fig. \ref{fig:eval_architecture}. Fig. \ref{fig:lineplots} shows an example lineplot through the generated and label image and their absolute difference. 

\red{Computation time is a limiting factor in image-guided interventions. Our unoptimized project-to-projection translation is capable of processing of $\sim24$ frames per second on a Nvidia Tesla V100 gpu. Naturally, this processing rate would be subject to additional delay or latency caused by the acquisition and preprocessing steps on the scanner. Nevertheless, the achievable processing time is fast enough to cope with common clinical X-ray fluoroscopy frame rates and will likely exceed the expected acquisition time for cone-beam MR projection images on a real scanner by far. 
}

% Lineplot images
\section{Discussion}
\label{sec:discussion}
% Detail markup images
\begin{figure*}[t]
	\includegraphics[width=1\textwidth]{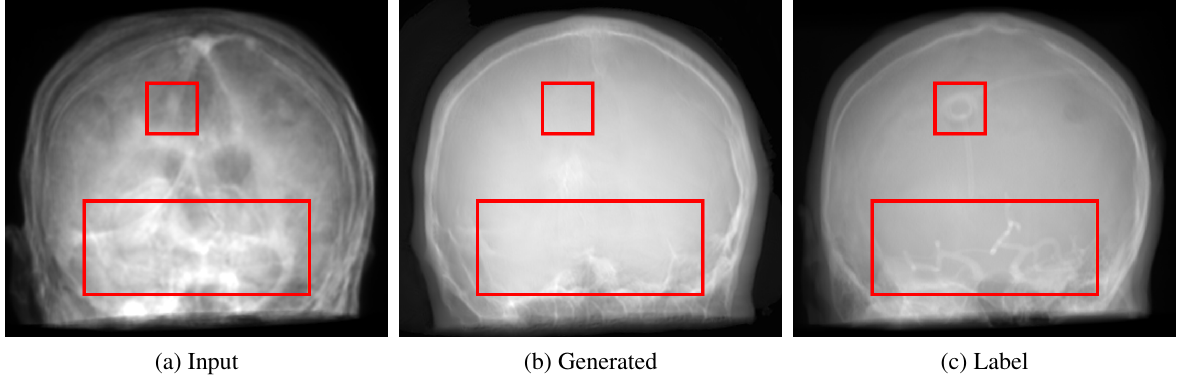}
	\caption{An example of missing information in the generated X-ray projection images. Details that are unobservable in the input MRI projections can, naturally, not be translated in the resulting generated projection images. \red{The small rectangle in the label image (c) outlines the entry point of a ventricular shunt while the larger rectangle frames contrasted cerebral arteries.} 
		While the devices were not present at the time of the acquisition of the underlying images, it is likely that none of these details will be captured by the MR imaging protocol.}
	\label{fig:markup}	
\end{figure*}
Quantitative comparison with other methods is difficult to conduct at this point. Numerous methods for image synthesis in medical imaging were proposed up to now, however, these exclusively target tomographic images. In contrast, projection images are significantly different in their general look regarding contrast, edges and superimpositions. 

A direct comparison between corresponding evaluation metrics would, therefore, be not meaningful for either side. 

\paragraph{Data generation}

\label{disc:data_gen} While the amount of patient scans that were available for this tasks is limited, our method of augmenting the data set by using multiple angles for forward projection is powerful. Data augmentation itself was proven remarkably beneficial in prior work~\cite{Perez2017}. Thereby, translation, rotation or simple warping are only the baseline and represent modified versions of the original image. In contrast, a forward projection is an X-ray  transform, i.e., an integral transformation. 
For a function $f(\vect{L})$ and the line $\vect{L}$, the X-ray transform $Xf(\vect{L})$  is the mapping % $F:\mathbb{R}^2\rightarrow\mathbb{R}$, calculated by 
\begin{equation}
Xf(\vect{L})= \int_{\mathbb{R}}^{}f(\vect{x}_0 + t\vect{\theta})dt \; ,
\end{equation}
where the unit vector $\vect{\theta}$ denotes the direction of the line $\vect{L}$ and $\vect{x}_0 $ is the starting point on the line.
If $f(\vect{L})$ is a constant function, i.e., one image, the resulting X-ray-transformed function is different for every single $\vect{\theta}$. Thereby, the previously mentioned popular augmentation methods are intrinsically part of the X-ray transform due to effects like perspective distortion, the projection trajectory, and more. 
What is, unfortunately, not solved by this data augmentation process are different medical indications of the patients. 
These come in various different appearances and shapes, e.g., only considering the cranial region, aneurysms, lesions, fractures, and many more. As the underlying dataset is composed of scans from different points in time, structural differences occur (see Fig. \ref{fig:markup}).  
Like with other learning-based algorithms, handling of these examples can only be covered by including them in the training database. It is, therefore, important to keep this fact in mind when dealing with data-based systems. Furthermore, multiple patient datasets are truncated in axial direction (see Fig. \ref{fig:projection_results}(c)) due to incomplete acquisition of the head. Nevertheless, our approach was able to deal with this challenge and produce the according truncated or untruncated synthesized projections. 

\paragraph{Architecture \& Objective Function}
As seen in Fig. \ref{fig:eval_architecture}(a) and (c), the redistribution of the residual blocks improves the visual impression of the generated images. 
The projections generated with the residual blocks at the high resolution levels resemble real head projections more closely. 
According to the motivation of this change described in Section \ref{sec:methods_architecture}, we conclude that this improvement is caused by the increased network capacity at the high-resolution layers. 
The consequent reduction in capacity in the coarser layers does not lead to a deterioration of the results. 
This may be due to the fact that the rough outline of the structures to be generated, i.e., the head, is the same in the MR and X-ray projections. 
Observation of the training process also showed that the general arrangement was already fixed after the first few epochs. 

High-frequency details are most of the time the most important part in X-ray projection imaging. Yet, most networks used for image synthesis are built in an encoder-decoder fashion which accumulates large portions of the available network capacity at the lower resolution layers to allow for an increase in filter dimension. Considering recent trends in super-resolution~\cite{Fan2018}, the usage of sub-pixel convolutions~\cite{Shi2016} might be of interest for future work. This technique performs the image generation almost completely on lower resolution levels without sacrificing on high-frequency details by reordering multiple low-dimensional feature maps into one high-dimensional image. Given the high memory requirements of image synthesis, this could be a valuable tool. 

Yet, only redistributing the network's capacity to higher resolution levels is not enough to guide the generation process in the desired direction. 
A more pronounced improvement can be observed by addition of the edge-weighting to the objective function. In Fig. \ref{fig:eval_architecture}(d), the fine details, i.e., edges, are much sharper compared to its counterpart (Fig.~\ref{fig:eval_architecture}(c)) which was not trained using this edge-weighting. Additionally, the amount of erroneously synthesized edges and borders is decreased when applying this weighting scheme. This improvement is not surprising when considering that only $\sim$9.8\,\% of the total image points belong to an edge according to our computed edge map. If the loss was not weighted, the error resulting from this small portion of image points would simply be overshadowed by the vast majority of low-frequency image points in the images.  

Another potentially interesting complement to our approach could be the inclusion of the temporal dimension. 
Fluoroscopic images consist usually of a sequence of images obtained from the same view. 
Similarly, while acquiring 3D tomographic scans, successive projections from slightly changed positions are recorded. 
In both cases, there is a certain degree of consistency between the successive projections. This can potentially be exploited by simultaneous processing of several projections in order to transfer this consistency to the results~\cite{Engelhardt, Lai_2018_ECCV, Aichert2015}. 

Despite the proposed additions and the resulting improvements, the projection-to-projection translation is not yet perfect. The high number of fine structures that are integrated on to the detector during the projection and thus overlap represent a great challenge for the synthesis. Combined with the initially discussed fact that the acquired signal is partly ambiguous, e.g., air and bone map to similar regions in most MR imaging sequences, providing a conclusive solution is a hard task. 
However, many interventional tasks do not require a perfect one-to-one correspondence. For example, smooth changes in the intensities may not be of great interest, assuming they are even perceivable for the human observer in the first place. 
In fact, an exact match of the images is not always the desired result. 
To apply many post-processing methods, subtle differences are needed to make these techniques possible. 
Tasks like one-shot digital subtraction angiography become feasible with the underlying synthesized and real contrasted X-ray projection images if slight differences between the projections are present. 

For future work, we hope to gain access to further corresponding data sets, especially those of body parts with more diverse information contained in the respective modalities. For example the chest features a greater diversity between the modalities with the X-ray based modality containing the ribs and spine, while all soft-tissue structures like the heart are better observable in the MR images. Additionally, we would like to evaluate the high-frequency component weighting also for other tasks that rely on accurate synthesis of structural details besides projection-to-projection translation. 

\section{Conclusion}
The underlying work presents an approach to synthesize X-ray projection images based on corresponding MR projections, which is potentially advantageous for hybrid MR and CT imaging applications. In doing so, we have identified distinct differences between the synthesis of projection images and volumetric images.  
In projection images, usually only edges and similar high-frequency structures contain important information. 
In order to emphasize this in the translation, we proposed an increase of the network capacity at higher resolution layers. In addition, a weighting of the high-frequency components in the image was introduced when calculating the error in the training process.
The proposed modifications and extensions proved to be valuable and showed clear improvements in the quantitative metrics with a decrease in the deviation from the ground truth by $\sim$25\,\%.  
In addition, particularly using the proposed weighting of high-frequency components in the training process resulted in a superior qualitative image impression. These results proved to yield sharper edges and fewer erroneously generated fine details. 
The advancing development in the fields of interventional MR and hybrid imaging motivates increased effort to solve the remaining problems of ambiguous signal and overlapping structures which still represent a challenging problem.

%\noindent LaTeX formats citations and references automatically using the bibliography records in your .bib file, which you can edit via the project menu. Use the cite command for an inline citation, e.g.  \cite{Hao:gidmaps:2014}.

%For data citations of datasets uploaded to e.g. \emph{figshare}, please use the \verb|howpublished| option in the bib entry to specify the platform and the link, as in the \verb|Hao:gidmaps:2014| example in the sample bibliography file.

\section*{Acknowledgements}
This work has been supported by the project P3-Stroke, an EIT Health innovation project. EIT Health is supported by EIT, a body of the European Union. Additional financial support for this project was granted by the Emerging Fields Initiative (EFI) of the Friedrich-Alexander University Erlangen-N\"urnberg (FAU) as well as Siemens Healthineers. Furthermore, we thank the NVIDIA Corporation for their hardware donation. 

\section*{Author contributions statement}
Conceptualization,~B.S.,~C.S.,~T.W.,~K.B.; Data curation,~P.H.~and~A.D.; Funding~acquisition,~A.D.~and~A.M.; Methodology,~B.S.; Validation,~B.S.; Writing—original~draft,~B.S.; Writing—review~and~editing,~B.S.,~C.S.,~T.W.,~K.B.,~P.H.,~A.D.,~A.M.

\section*{Additional information}
\textbf{Competing interests}: The authors declare no competing interests.\\% 
\textbf{Data availability}: For legal reasons, the authors are not able to share the underlying clinical data sets. 
\\
%
%The corresponding author is responsible for submitting a \href{http://www.nature.com/srep/policies/index.html#competing}{competing interests statement} on behalf of all authors of the paper. This statement must be included in the submitted article file.

\end{document}